\documentclass[preprint,12pt]{elsarticle}
\usepackage{booktabs} % three-line tables
\usepackage{amssymb}
\usepackage{amsmath}
\usepackage{geometry}
\usepackage{xcolor}
\usepackage[normalem]{ulem}

\newcommand{\stkout}[1]{\ifmmode\text{\sout{\ensuremath{#1}}}\else\sout{#1}\fi}
\newcommand{\add}[1]{#1}
\newcommand{\del}[1]{}
\newcommand{\change}[2]{\del{#1}\add{#2}}

\usepackage{float} % allow [h] float placement
\journal{}

\DeclareUnicodeCharacter{03B1}{\ensuremath{\alpha}}
\begin{document}
\sloppy

\begin{frontmatter}

\title{Study of ordering in (MoCrTi)$_{100-x}$Al$_x$ refractory high-entropy alloys using machine learning interatomic potential}

\author[leoben]{Jiyao Zhang}
\author[leoben]{Klemens Lechner}
\author[graz]{Markus Maßwohl}
\author[graz]{Petra Spoerk-Erdely}
\author[leoben]{David Holec}

\affiliation[leoben]{organization={Department of Materials Science, Montanuniversität Leoben},
            addressline={Franz-Josef-Strasse 18}, 
            city={Leoben},
            postcode={8700}, 
            country={Austria}}

\affiliation[graz]{organization={Institute of Materials Science, Joining and Forming, Graz University of Technology},
            addressline={Kopernikusgasse 24/I}, 
            city={Graz},
            postcode={8010}, 
            country={Austria}}

\begin{abstract}
\change{Refractory high-entropy alloys have emerged as promising candidates for high-temperature applications due to their exceptional mechanical properties. Understanding the thermodynamic mechanisms underlying chemical ordering in these complex systems is critical for optimizing their performance. In this work, by utilizing a universal machine learning interatomic potential with hybrid Monte Carlo and molecular dynamics simulations, the temperature-dependent thermodynamics and mechanical properties of (MoCrTi)$_{100-x}$Al$_x$ system have been investigated. The heat capacities and short-range order parameters reveal distinct order-disorder transition behaviors. While the Mo$_{25}$Cr$_{25}$Ti$_{25}$Al$_{25}$ and Mo$_{32}$Cr$_{32}$Ti$_{32}$Al$_{4}$ alloys exhibit a single transition dominated by the synergistic ordering of B2-type atomic pairs, the Mo$_{28}$Cr$_{28}$Ti$_{28}$Al$_{16}$ and Mo$_{30}$Cr$_{30}$Ti$_{30}$Al$_{10}$ alloys display two separate transitions: a low-temperature stage driven by specific pairs (Mo--Al in Mo$_{28}$Cr$_{28}$Ti$_{28}$Al$_{16}$; Al--Al in Mo$_{30}$Cr$_{30}$Ti$_{30}$Al$_{10}$) and a high-temperature stage governed by the remaining pairs. Structural analysis indicates that in the low-temperature ordered B2 ordering, Mo and Al share one sublattice while Cr and Ti share the other. Furthermore, the relationship between ordering and mechanical stiffness has been identified. Ordering significantly enhances the elastic constants and moduli, and gives rise to a non-monotonic compositional dependence. Unlike random solid solutions, where stiffness increases monotonically with decreasing Al content, ordered configurations exhibit a non-monotonic trend, peaking at the Mo$_{30}$Cr$_{30}$Ti$_{30}$Al$_{10}$ alloy. This enhancement is attributed to an optimized population of stiff atomic pairs induced by strong short-range order. These findings provide fundamental insights into the interplay between compositions, chemical ordering, and mechanical performance, offering guidance for the design of refractory high-entropy alloys.}{Refractory high-entropy alloys are promising candidates for high-temperature applications, yet the effects of composition on their chemical-ordering pathways and mechanical properties remain insufficiently understood. Here, a universal machine-learning interatomic potential combined with Monte Carlo and molecular dynamics simulations is employed to investigate the temperature-dependent thermodynamic and mechanical behavior of (MoCrTi)$_{100-x}$Al$_x$ alloys. Atomic configurations, sublattice occupations, and simulated diffraction intensities reveal pronounced B2-type chemical ordering at low temperatures, with Mo and Al preferentially occupying one sublattice and Cr and Ti occupying the other. The configurational heat capacities and short range-order parameters further reveal a strong composition dependence of the ordering pathway. The Mo$_{25}$Cr$_{25}$Ti$_{25}$Al$_{25}$ and Mo$_{32}$Cr$_{32}$Ti$_{32}$Al$_4$ alloys exhibit a single dominant ordering stage involving cooperative changes in multiple B2-type pair correlations. By contrast, Mo$_{28}$Cr$_{28}$Ti$_{28}$Al$_{16}$ and Mo$_{30}$Cr$_{30}$Ti$_{30}$Al$_{10}$ exhibit two distinct ordering stages. Their low-temperature features are associated primarily with changes in the Mo--Al and Al--Al correlations, respectively, whereas their high-temperature features involve collective changes in the remaining B2-type pair correlations. Chemical ordering also fundamentally alters the composition dependence of mechanical stiffness. Whereas the elastic constants and moduli of chemically disordered configurations increase approximately monotonically with decreasing Al content, those of the ordered configurations exhibit a non-monotonic dependence and reach a maximum in Mo$_{30}$Cr$_{30}$Ti$_{30}$Al$_{10}$. This anomalous enhancement originates from a short range order induced redistribution of atomic pairs that increases the population of mechanically stiff bonds, particularly Mo--Cr pairs. These results establish a direct atomistic connection among alloy composition, multistage chemical ordering, and mechanical stiffness, providing guidance for tuning the mechanical behavior of RHEAs through compositional control.}

\end{abstract}

% %%Graphical abstract
% \begin{graphicalabstract}
% %\includegraphics{grabs}
% \end{graphicalabstract}

% %%Research highlights
% \begin{highlights}
% \item Research highlight 1
% \item Research highlight 2
% \end{highlights}

%% Keywords

\end{frontmatter}

\section{Introduction}

The aerospace industry increasingly demands refractory alloys capable of withstanding extreme thermal environments. Refractory high entropy alloys (RHEAs), composed of principal elements with high melting points, have emerged as promising candidates to meet these challenges. Since the initial report of NbMoTaW(V) RHEAs by Senkov et al.~\cite{Senkov2010-ym, Senkov2011-dj} in the 2010s, RHEAs have attracted significant research attention~\cite{Xiong2023-fx}. Notably, RHEAs exhibit high-temperature mechanical performance superior to that of traditional Ni-based superalloys~\cite{Diao2017-px}, positioning them as ideal materials for next-generation elevated temperature applications. Early RHEAs' development focused on single-phase solid solutions. Systems such as ZrNbTiVHf~\cite{Feuerbacher2018-nd} and HfNbTiZr~\cite{Wu2014-zt} typically form stable single-phase body-centered cubic (BCC, Strukturbericht designation A2, space group $Im\bar3m$) microstructures due to the high mutual solubility of their constituent refractory metal elements. \add{However, the addition of specific elements, particularly Al, can promote the formation of multiphase microstructures—including Laves and A15 phases—as well as pronounced short-range order (SRO) in RHEAs, thereby enhancing their mechanical properties.}\del{in Ta/Nb--MoCrTiAl systems. While complex phases can introduce enhancement as well as brittleness, controlled precipitation strengthening has shown great potential.} For instance, Yang et al.~\cite{Yang2025-to} demonstrated that the precipitation-strengthened Ta$_{27.3}$Mo$_{27.3}$Ti$_{27.3}$Cr$_{8}$Al$_{10}$ (at.\%) alloy, characterized by a disordered A2 matrix and ordered B2 (space group $Pm\bar3m$) precipitates, exhibits substantially higher creep resistance (minimum creep rate reported four orders of magnitude lower) compared to other single-phase A2 TiZrHfNbTa or B2 NbMoCrTiAl RHEAs. \add{Qi et al.~\cite{Qi2024-jr} reported aluminum-enriched B2 HEAs, the Al$_{15}$Hf$_{25}$Nb$_{32}$Ti$_{28}$ and Al$_{15}$Hf$_{25}$Nb$_{32}$Ti$_{28}$ (at.\%) exhibited comparable strength but with substantially improved ductility compared with other Al-containing RHEAs.} Nevertheless, precise phase control remains a challenge. Müller et al.~\cite{Muller2020-rt} reported the formation of brittle Laves phases (e.g., Cr$_2$Nb, Strukturbericht designation C15, space group $Fd\bar3m$) and A15 (space group $Pm\bar3n$) phases in Nb/Ta--MoCrTiAl systems, noting that suppressing these detrimental phases requires reducing Cr and Ta/Nb contents. \change{In this context, the MoCrTiAl system presents a distinct advantage.}{Accordingly, unlike the multiphase RHEAs reported in previous studies, the MoCrTiAl system exhibits a simple A2 + B2 microstructure~\cite{Laube2024-bw, Laube2020-zy}. This structural simplicity provides a distinct advantage for investigating short-range ordering.} Laube et al.~\cite{Laube2024-bw, Laube2020-zy} found that this system can suppress Laves phase formation while retaining B2 phase\del{which is beneficial to the high-temperature performance}. By varying the Al content, the volume fraction of B2 phase in the A2 matrix can be tailored to optimize mechanical properties: increasing Al typically raises the B2 phase fraction and thus improves strength, but it often reduces ductility~\cite{Laube2024-bw, Laube2020-zy}. Consequently, by achieving a balanced, coherent B2/A2 microstructure (often referred to as a superalloy-like structure), it is possible to optimize the strength-ductility trade-off.

However, a detailed understanding of the underlying atomic-scale mechanisms governing these order-disorder transitions and configuration stability is missing. The multi-principal elements nature of RHEAs makes theoretical modeling particularly challenging. Traditional approaches, such as cluster expansion and the embedded atom method, are prone to overfitting~\cite{Liu2023-kl}. Furthermore, first-principles Density Functional Theory (DFT) calculations are computationally prohibitive for large-scale RHEAs simulations due to the substantial resources required~\cite{Liu2023-kl}. The past decades have brought great progress in machine learning (ML), which has encouraged its application in materials science. For instance, Li et al.~\cite{Li2020-ic} developed a machine learning interatomic potential using the spectral neighbor analysis potential (SNAP) approach to study the complex strengthening mechanisms in the NbMoTaW alloy. This potential has also been utilized to investigate the hierarchical energy landscape of screw dislocation motion in RHEAs~\cite{Wang2022-rv}. Similarly, Gubaev et al.~\cite{Gubaev2023-tl} developed two ML-based potentials, a moment tensor potential (MTP) and a Gaussian moment neural network (GM-NN), which accurately capture atomic interactions in the TaVCrW system. While these ML-based models are significantly more accurate than conventional embedded-atom method (EAM) potentials (also fitted in Ref.~\cite{Gubaev2023-tl}), they are typically developed for specific alloy systems and require extensive DFT training datasets (e.g., over 6000 configurations~\cite{Gubaev2023-tl}), limiting their transferability. Fortunately, the rapid advancement of data-driven materials science and large-scale databases, such as the Materials Project~\cite{Jain2013-ch}, has facilitated the development of universal Machine Learning Interatomic Potentials (uMLIPs)~\cite{Shuang2025-dy}. These universal models offer a promising solution to HEA modeling challenges by effectively learning from ab-initio data across the periodic table, enabling efficient simulations of chemically complex compositions~\cite{Xia2025-xo}.

In this research, we employ a combination of atomistic modeling techniques, including molecular dynamics (MD), Monte Carlo (MC), and DFT, to investigate the SRO in (MoCrTi)$_{100-x}$Al$_x$ RHEAs as a function of composition and temperature. First, we detail the atomistic simulation protocols, \change{SRO}{chemical ordering} characterization methods, and \add{configurational} heat capacity calculations. Next, we analyze the \add{configurational} heat capacity curves and investigate the correlation between SRO parameters and the order/disorder transition. Finally, the impact of SRO evolution on the mechanical properties of the RHEAs is determined.

\section{Computational methods}
\subsection{Hybrid Monte Carlo/molecular dynamics simulations}

Hybrid MC/MD simulations were conducted using the Atomic Simulation Environment (ASE)~\cite{Hjorth-Larsen2017-zy} combined with the uMLIP GRACE-1L-OAM model developed by Bochkarev et al.~\cite{Bochkarev2024-qj}. The GRACE-1L-OAM model offers an optimal balance between accuracy and computational efficiency~\cite{Shuang2025-dy}, and GRACE uMLIP's reliability has been extensively validated in prior studies~\cite{Shuang2025-dy, Reiners-Sakic2026-pq}. To systematically investigate the influence of aluminum content on the order-disorder transition, four alloy compositions with decreasing Al concentrations were selected:  Mo$_{25}$Cr$_{25}$Ti$_{25}$Al$_{25}$, Mo$_{28}$Cr$_{28}$Ti$_{28}$Al$_{16}$, Mo$_{30}$Cr$_{30}$Ti$_{30}$Al$_{10}$, and Mo$_{32}$Cr$_{32}$Ti$_{32}$Al$_{4}$ (at.\%). For brevity, these alloys are hereafter referred to as Al25, Al16, Al10, and Al4, respectively.

\begin{figure}[htbp]
    \centering
    \includegraphics[width=1.0\linewidth]{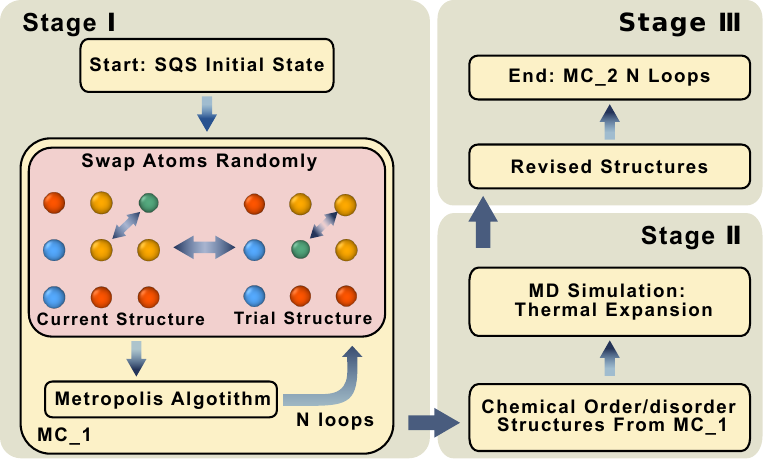}
    \caption{The workflow of hybrid MC/MD simulations. For a detailed explanation of individual stages, see the text.}
    \label{fig:combined MC_MD Algorithm}
\end{figure}

Initial atomic configurations were generated as special quasi-random structures (SQS) using the sqsgenerator code~\cite{Gehringer2023-mr}, based on a BCC $4 \times 5 \times 5$ supercell containing 200 atoms with lattice constant $a = 3.17\,\text{\AA}$. For all compositions, the simulation protocol at each temperature (i.e. 200, 400, 600, 800, 850, 900, 1000, 1050, 1100, 1150, 1200, 1250, 1300, 1400, 1600, 1800, and 2000\,K) comprised three sequential stages, as illustrated in Fig.~\ref{fig:combined MC_MD Algorithm}:

\begin{enumerate}
    \item \textbf{Stage I: MC Equilibration}: Starting from the initial SQS configuration, trial structures were generated by randomly  swapping two atoms while keeping the lattice fixed. The Metropolis algorithm~\cite{Metropolis1953-de} was employed to determine the acceptance of each new configuration. Swaps between atoms of the same element were skipped to ensure that each attempt altered the configuration. To ensure thermodynamic robustness, the equilibration was performed for 2,000,000 MC steps.

    \item \textbf{Stage II: MD Thermal Expansion}: The equilibrated structures at each temperature for all the compositions obtained from Stage I served as the starting configurations for $NPT$ (obaric-isothermal) MD simulations. In this stage, the cell volume was allowed to relax while the cell shape was constrained to allow for thermal expansion. The simulations were conducted for 50,000 steps with a time step of 2.0\,fs (total duration of 100\,ps).

    \item \textbf{Stage III: Production MC Simulation}: Starting from the thermally expanded structures, additional 2,000,000 MC steps were performed to collect sufficiently thermally equilibrated data for the subsequent calculation of configurational heat capacity and SRO parameters. 
\end{enumerate}

Validation data for the first MC and MD simulations are provided in the Supplementary Material. As shown in Fig.~S1, all the potential energies stabilize after approximately 1,000,000 steps, confirming that the systems reached thermodynamic equilibrium for the first MC simulations (performed on a fixed lattice). The presence of SRO is evidenced by the negative SRO parameters observed in Fig.~S2. These equilibrated configurations served as starting points for the subsequent MD thermal expansion calculations. The MD results demonstrate thermal equilibration through the convergence of lattice constants (Fig.~S3) and exhibit a linear expansion trend (Fig.~S4). Results from the second MC simulations will be discussed later.

In MC simulations, the constant-volume \add{configurational} heat capacities (\change{$C_V$}{$C_{V,conf}$}) are calculated based on energy fluctuations within the $NVT$ ensemble (constant  number  of  atoms,  volume,  and  temperature), a fundamental concept in statistical mechanics~\cite{Woodgate2023-fn, Allen2017-rd}. \add{The core formula is:
\begin{equation}
C_{V,conf}(T) = \frac{ \langle E^2 \rangle - \langle E \rangle^2 }{ k_B T^2 }\ ,
\label{eq:heat_capacity}
\end{equation}}
where $\langle E \rangle$ is the ensemble average of the global potential energy, $\langle E^2 \rangle$ is the ensemble average of the square of the global potential energy, $k_B$ is the Boltzmann constant $(8.617 \cdot 10^{-5}\,\text{eV/K})$, and $T$ is the absolute temperature.
   
To quantitatively characterize the degree of SRO within the alloys, the Warren-Cowley parameters were employed. The SRO parameter $\alpha_{\xi\eta}^{i}(\sigma)$, representing a specific atomic pair within a given coordination shell, is defined as follows~\cite{Gehringer2023-tz}: 
\begin{equation}
\alpha_{\xi\eta}^{i}(\sigma) = 1 - \frac{N_{\xi\eta}^i(\sigma)}{N \, M^i\, x_\xi \, x_\eta}\ ,
\label{eq:sro_definition}
\end{equation}
where $\xi$ and $\eta$ denote atomic species in the system, $N_{\xi\eta}^i(\sigma)$ is the actual observed number of $\xi$--$\eta$ atomic pairs in the $i^{\mathrm{th}}$ coordination shell for configuration $\sigma$, $N$ is the number of lattice sites, $M^i$ is the number of nearest neighbors in the given atomic arrangement, and $x_\xi$ and $x_\eta$ are the concentrations of species $\xi$ and $\eta$, respectively. $N\, M^i \, x_\xi \, x_\eta$ represents the expected number of $\xi$--$\eta$ pairs in a perfectly random solid solution (ideal mixture). $\alpha = 0$ signifies that atoms are distributed randomly (ideal solid solution). When $\alpha > 0$, there are fewer $\xi$--$\eta$ pairs than would be expected in a random distribution, indicating segregation (repulsion between species). Conversely, when $\alpha < 0$, more $\xi$--$\eta$ pairs exist than predicted by \add{a} random distribution, suggesting ordering (attraction between species).

\add{To further characterize the B2-type chemical ordering, the long-range order (LRO) parameter was calculated, and the diffraction patterns were simulated using VESTA~\cite{Momma2011-ii}. The B2 lattice can be divided into two sublattices, denoted as $\alpha$ and $\beta$, which together form the A2 lattice. These two sublattices correspond to ``corner'' and ``center'' positions in Fig.~\ref{fig:b2_congigurations}. In a disordered A2 configuration, each element is randomly distributed between the two sublattices, whereas B2-type ordering is characterized by preferential occupation of one sublattice. The element-specific LRO parameter $\eta_i$ is defined as:
\begin{equation}
\eta_i = \frac{X_{i-\alpha} - X_{i-\beta}}{X_{i-\alpha} + X_{i-\beta}}\ ,
\label{eq:LRO_definition}
\end{equation}
where $X_{i-\alpha}$ and $X_{i-\beta}$ denote the numbers of atoms of element $i$ occupying the $\alpha$ and $\beta$ sublattices, respectively. The parameter $\eta_i$ ranges from $-1$ to $1$. A value of $\eta_i \approx 0$ indicates no preferential occupation of either sublattice, whereas $\eta_i=1$ and $\eta_i=-1$ correspond to exclusive occupation of the $\alpha$ and $\beta$ sublattices, respectively, representing the maximum degree of element-specific sublattice ordering.}

\subsection{DFT \change{S}{s}imulation\add{s}}
    
\begin{figure}[htbp]
    \centering
    \includegraphics[width=0.3\linewidth]{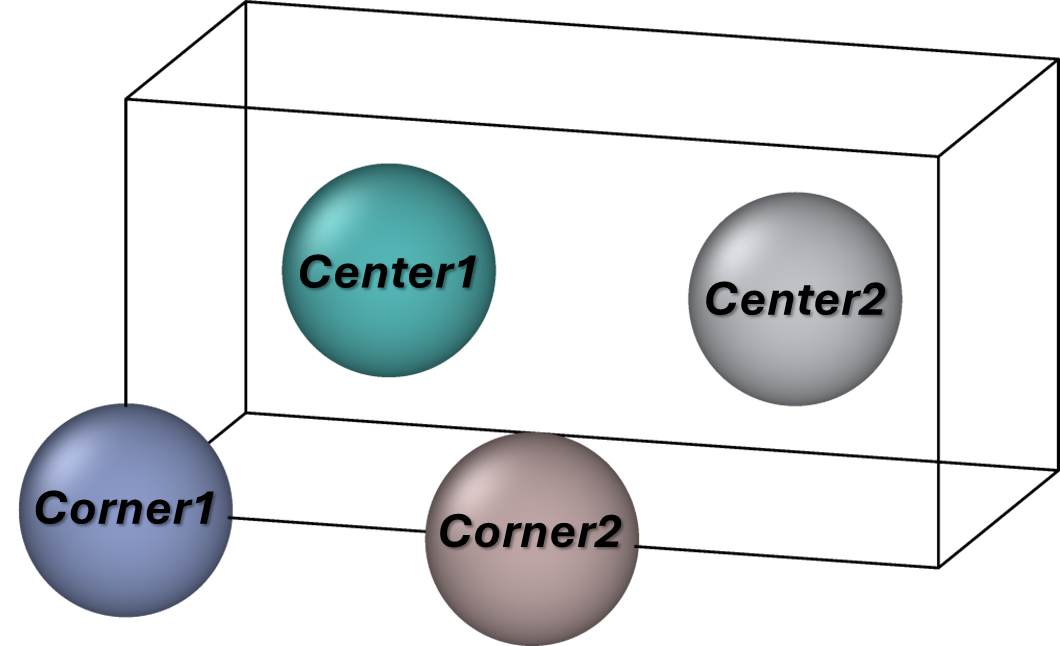}
    \caption{Schematic of the B2-based configuration.}
    \label{fig:b2_congigurations}
\end{figure}

The quantum-mechanical calculations were conducted utilizing the Vienna Ab-initio Simulation Package (VASP)~\cite{Kresse1996-wp,Kresse1996-zu} within the DFT~\cite{Hohenberg1964-hc,Kohn1965-bz} framework, employing the projector augmented-wave (PAW) pseudopotentials~\cite{Kresse1999-xx} and the Perdew-Burke-Ernzerhof (PBE) exchange-correlation functional~\cite{Perdew1996-rj}.
The $2\times1\times1$ supercells of the conventional BCC unit cell contain 4 atoms, which split into two sublattices resembling the corresponding B2 structure with two sublattices: one containing the corner positions and one containing the central positions. By placing Mo, Cr, Ti, and Al on these different sublattices, various configurations can be obtained, as illustrated in Fig.~\ref{fig:b2_congigurations}. Depending on the occupation of sublattices, they are named as shown in Table~\ref{tab:configurations_names}. A plane-wave cutoff energy (\texttt{ENCUT}) of $500\,\text{eV}$ and a Gamma-centered $k$-point mesh of $5\times10\times10$ were employed. The convergence criteria for the electronic self-consistent step and the ionic relaxation were set to $10^{-5}\,\text{eV}$ and $10^{-3}\,\text{eV/\AA}$, respectively. Additionally, $\texttt{ISIF} = 8$ was used to exclude the influence of cell shape relaxations.

\begin{table}[htbp]
    \caption{Various configurations of the supercell with different atomic arrangements.}
    \centering
    \begin{tabular}{ccccc}
        \toprule
        Configuration & Corner 1 & Corner 2 & Center 1 & Center 2 \\
        \midrule
        (Mo,Cr)\_(Ti,Al) & Mo & Cr & Ti & Al\\
        (Mo,Ti)\_(Cr,Al) & Mo & Ti & Cr & Al\\
        (Mo,Al)\_(Cr,Ti) & Mo & Al & Cr & Ti\\
        \bottomrule
    \end{tabular}
    \label{tab:configurations_names}
\end{table}

\subsection{Mechanical properties}

The mechanical properties, including elastic constants and elastic moduli, were obtained by averaging the results from the last six equilibrated MC snapshots. For cubic crystals, there are three independent elastic constants, namely $C_{11}$, $C_{12}$, and $C_{44}$~\cite{Nye1985-wr}. These elastic constants were calculated using the uMLIP-based stress--strain method with universal linear-independent coupling strains applied to the relaxed equilibrium structures, yielding the elastic stiffness matrix $C_{ij}$ according to~\cite{Yu2010-hs}. \add{Subsequently, the obtained $C_{ij}$ tensor was projected onto cubic symmetry according to the methodology of Moarker and Norris~\cite{Moakher2006-zm}}. The maximum strain amplitude was set to 0.5\%. The elastic moduli were calculated using the corresponding formulas~\cite{Abdoshahi2021-hm, Nye1985-wr, Hill1952-ix}. 

To evaluate the relative stiffness of different atomic pairs, the pair stiffness (force constant, $k$ [$\text{eV/\AA}^2$]) was determined from the curvature of the energy vs. bond-length relationship~\cite{Song1999-qn}. For each elemental pair among Mo, Cr, Ti, and Al, an ideal BCC unit cell was constructed with one
atom located at the corner sublattice and the other at the centered sublattice. For each pair, the lattice constant was varied from $2.70$ to
$3.60~\text{\AA}$, and the potential energy was calculated using the GRACE uMLIP. The bond length $r$ was
defined as the distance between the corner and central atoms. The resulting energy ($E(r)$) vs. bond-length curve was then fitted near
the equilibrium distance within $r_{\mathrm{min}} \pm 0.15\,\text{\AA}$ using a harmonic approximation:
\begin{equation}
E(r) \approx E_0 + \frac{1}{2}k_{\mathrm{cell}}(r-r_0)^2\ ,
\label{eq:harmonic_approximation}
\end{equation}
where $E_0$ and $r_0$ are the equilibrium energy and bond length, and cell stiffness $k_{\mathrm{cell}}$ is obtained from the curvature of the fitted energy curve. The pair stiffness was estimated as:
\begin{equation}
k = \frac{k_{\mathrm{cell}}}{8}\ ,
\label{eq:bond_stiffness}
\end{equation}
where the factor of 8 accounts for the eight equivalent nearest-neighbor bonds in the BCC unit cell. The obtained $k$ values were then used to compare the relative stiffness of different atomic pairs.

\section{Results and discussion}
\subsection{\add{Configurational} heat capacity\del{ of RHEAs}}
\begin{figure}[htbp]
    \centering
    \includegraphics[width=0.75\linewidth]{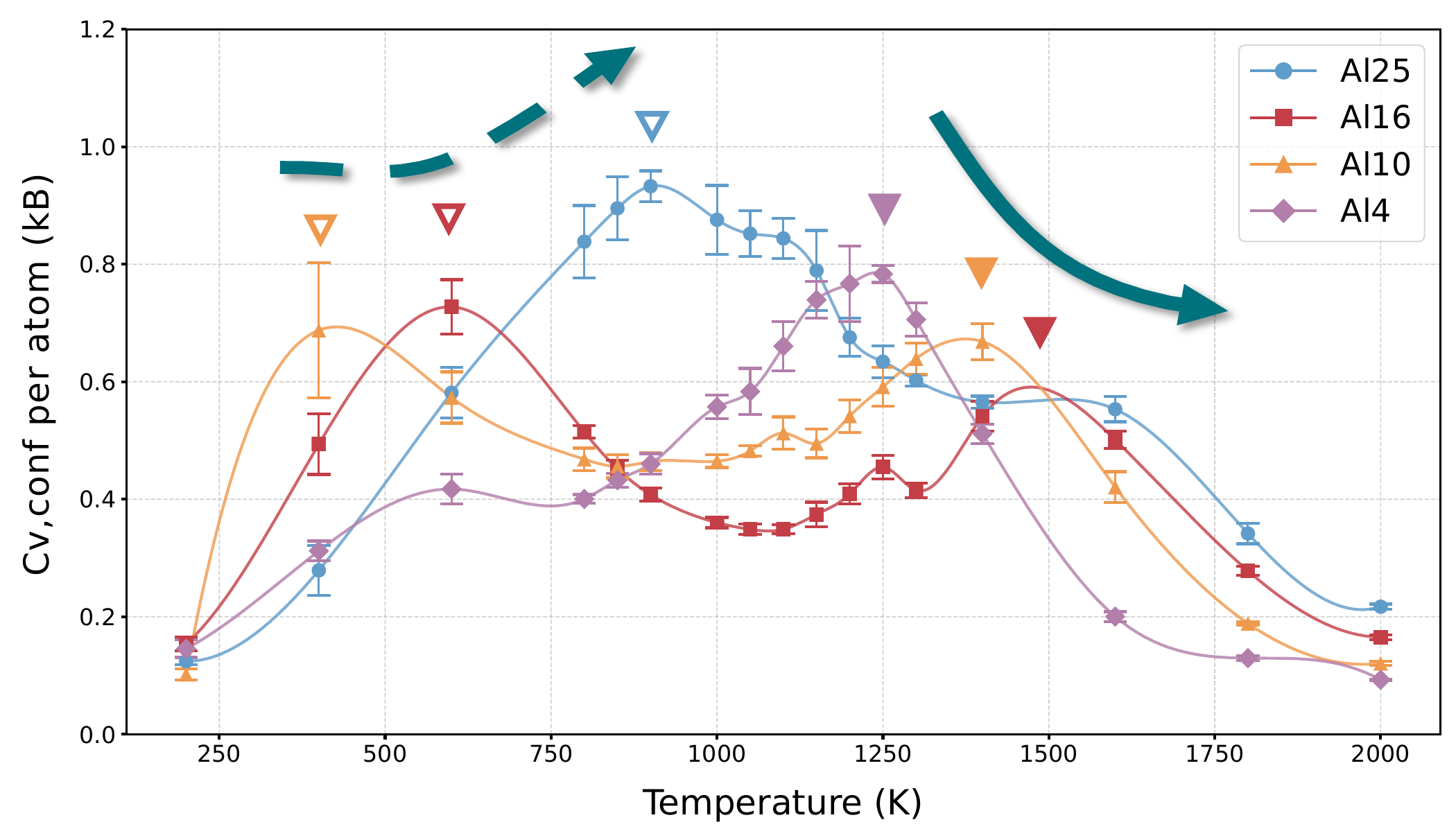}
    \caption{\add{Configurational heat capacity per atom versus temperature for four alloy compositions: Al25 (blue circles), Al16 (red squares), Al10 (orange triangles), and Al4 (purple rhombus). In the figure, hollow triangles and dashed arrows indicate the position and shift of the low-temperature peaks, respectively; solid triangles and solid arrows indicate the position and shift of the high-temperature peaks, respectively.}}
    \label{fig:Heat_capacities_of_alloys}
\end{figure}

The energy evolution during the second MC simulation is provided in the Supplementary Material. As shown in Fig.~S5, the cumulative average energy curves demonstrate that all systems reached a stable thermodynamic state after 1,000,000 steps. To ensure robustness of the results, only data collected beyond this point were used for the statistical analysis to determine the \add{configurational heat capacity per atom, $C_{V,conf}(T)/N$, as shown in Fig.~\ref{fig:Heat_capacities_of_alloys}. At each temperature, the data collected after 1,000,000 MC steps were divided into five blocks, and $C_{V,conf}$ was calculated separately for each block. The resulting mean values and error bars are shown in Fig.~\ref{fig:Heat_capacities_of_alloys}. The connecting curves were smoothed using cubic-spline interpolation.}

The curves change significantly with the Al content varying. The \change{$C_V(T)$}{$C_{V,conf}(T)/N$} profiles of the Al25 (blue circles) and Al4 (purple rhombus) alloys are characterized by a single strong peak. In contrast, the intermediate compositions Al16 (red squares) and Al10 (orange triangles) exhibit bimodal behavior, featuring distinct peaks at both low and high temperatures. With increasing Al content, the magnitude of the high-temperature peak decreases, its shape flattens and broadens, and it shifts to higher temperatures before finally vanishing in the Al25 alloy (where it is seen only as a shoulder around $1500\,\text{K}$). Conversely, a low-temperature peak gradually emerges and systematically shifts towards higher temperatures (to the right) with increasing Al content, eventually evolving into the dominant feature of the equiatomic Al25 alloy. 

The \add{configurational} heat capacity of the alloy is derived from the fluctuations in potential energy. Consequently, peaks in the \add{configurational} heat capacity curves signify rapid potential energy changes driven by configurational evolution, specifically marking the order-disorder transition \add{are most pronounced}. Laube et al.~\cite{Laube2020-zy} analyzed dynamic differential scanning calorimetry (DSC) curves for MoCrTiAl systems and observed $\lambda$-shaped peaks, which indicate a second-order or continuous phase transformation in equiatomic MoCrTiAl as well as in MoCrTi--15Al and MoCrTi--10Al alloys. The experimental results indicate a clear trend: as the Al content increases, the $\lambda$-shaped peaks shift to higher temperatures~\cite{Laube2020-zy}. This is in good agreement with our simulation results, which similarly show that both the low-temperature and high-temperature peaks shift to higher temperatures with increasing Al concentration. Furthermore, Laube et al.~\cite{Laube2020-zy} reported a transition peak at around $1273\,\text{K}$ for the equiatomic MoCrTiAl alloy. Similarly, Chen et al.~\cite{Chen2019-nn} observed a transition peak at approximately $1263\,\text{K}$ for the same alloy. In contrast, the transition in our simulations occurs around $900\,\text{K}$, with a high-temperature shoulder around $1500\,\text{K}$. The discrepancy in the absolute transition temperatures may be attributed to the inherent limitations of the uMLIP utilized. Specifically, the training datasets of this GRACE-1L-OAM uMLIP (i.e. OMat24~\cite{Barros-Luque2026-sw}, Alexandria~\cite{Schmidt2023-cj}, and MPtrj~\cite{Deng2023-jn}), are predominantly composed of $0\,\text{K}$ DFT relaxations and lower-order compounds. The lack of high-temperature HEA configurations in the training data limits the potential accuracy in fully capturing the complex high-temperature thermodynamics of such RHEA systems~\cite{Barros-Luque2026-sw, Schmidt2023-cj, Deng2023-jn}. Furthermore, fixed-lattice MC simulations exclude local distortions, which are known to be essential for HEA properties. \add{To corroborate this estimate of the heat capacity, the phonon contribution was additionally evaluated using the Debye model~\cite{Bhandari2020-lp, Boucetta2014-km}. The resulting combined heat capacity (Fig.~S9 in Supplementary Material) exhibits the same transition trend and features as the configurational heat capacity shown in Fig.~\ref{fig:Heat_capacities_of_alloys}. Further details and results are provided in section 2 of the Supplementary Material entitled ``Debye model estimated heat capacity''.}

\subsection{\change{Short-range order parameters in RHEAs}{B2 ordering analysis}} 

\begin{figure}[htbp]
    \centering
    \includegraphics[width=1\linewidth]{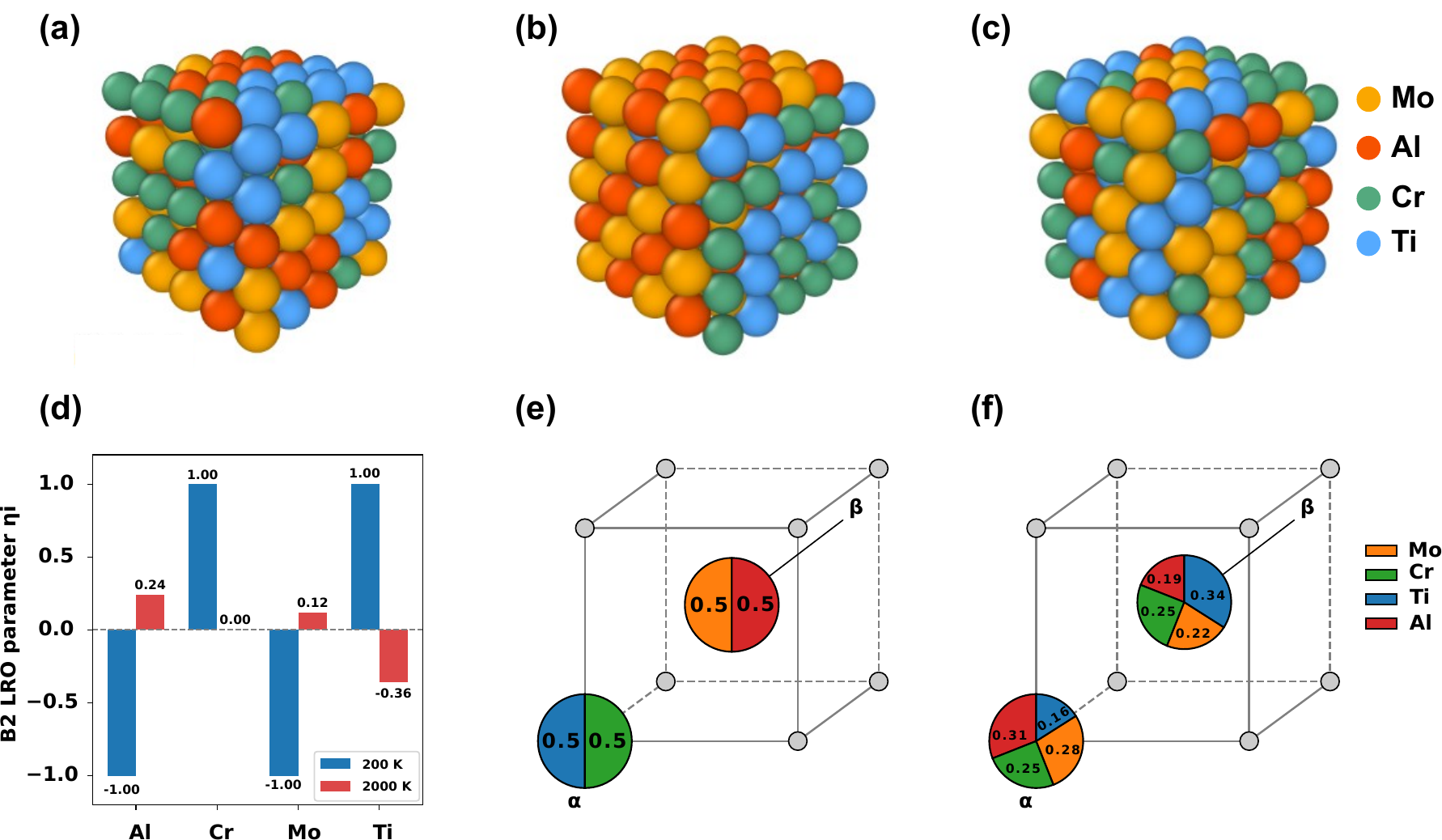}
    \caption{\add{Snapshots from MC simulations of the Al25 alloy: (a) initial SQS  state, (b) equilibrated ordered structure at $200\,\text{K}$, (c) equilibrated random structure at $2000\,\text{K}$, (d) corresponding LRO parameters at $200\,\text{K}$ and $2000\,\text{K}$, and schematic representations of the element-resolved occupations of the two B2 sublattices at (e) $200\,\text{K}$ and (f) $2000\,\text{K}$.}} 
    \label{fig:snapshots of structure}
\end{figure}

\change{Fig.~\ref{fig:snapshots of structure} shows}{Figs.~\ref{fig:snapshots of structure}(a--c) show} structural snapshots at different simulation stages to elucidate the ordering of the Al25 alloy. Fig.~\ref{fig:snapshots of structure}(a) displays the SQS used as the initial state of MC calculations. In contrast to this random initial state, a B2-like ordering is clearly visible in Fig.~\ref{fig:snapshots of structure}(b), representing the equilibrated state at $200\,\text{K}$. While the classical binary Ti-Al B2 structure consists of Ti and Al atoms occupying distinct sublattices~\cite{Holec2016-id}, the Al25 RHEA adopts a pseudo-binary configuration where (Mo, Al) atoms occupy one sublattice and (Cr, Ti) atoms occupy the other. Finally, the representative snapshot at $2000\,\text{K}$ confirms the random solid solution at high temperature (Fig.~\ref{fig:snapshots of structure}(c)). 

\add{This sublattice partitioning is quantitatively confirmed by the LRO parameters and element-resolved site occupancies in Figs.~\ref{fig:snapshots of structure}(d) and (e). At $200\,\text{K}$, the LRO parameter $\eta_i$ equals to $1$ or $-1$, indicating all the elements can be exclusively found on the $\alpha$ (Cr, Ti) or $\beta$ (Mo, Al) sublattices. The site occupation fractions of approximately 0.5 for each preferred elemental pair further demonstrate the pseudo-binary nature of the ordered state. In contrast, the equilibrated configuration at $2000\,\text{K}$ exhibits a nearly random solid-solution arrangement. Correspondingly, the magnitudes of all LRO parameters decrease substantially toward 0 (Fig.~\ref{fig:snapshots of structure}(d)), indicating the loss of long-range chemical order. The element-resolved occupancies in Fig.~\ref{fig:snapshots of structure}(f) likewise show that all four elements are distributed across both $\alpha$ and $\beta$ sublattices. These results demonstrate a temperature-driven order-to-disorder transition from a low-temperature B2-like ordering phase to a high-temperature disordered A2 solid solution.}

\begin{figure}[htbp]
    \centering
    \includegraphics[width=1\linewidth]{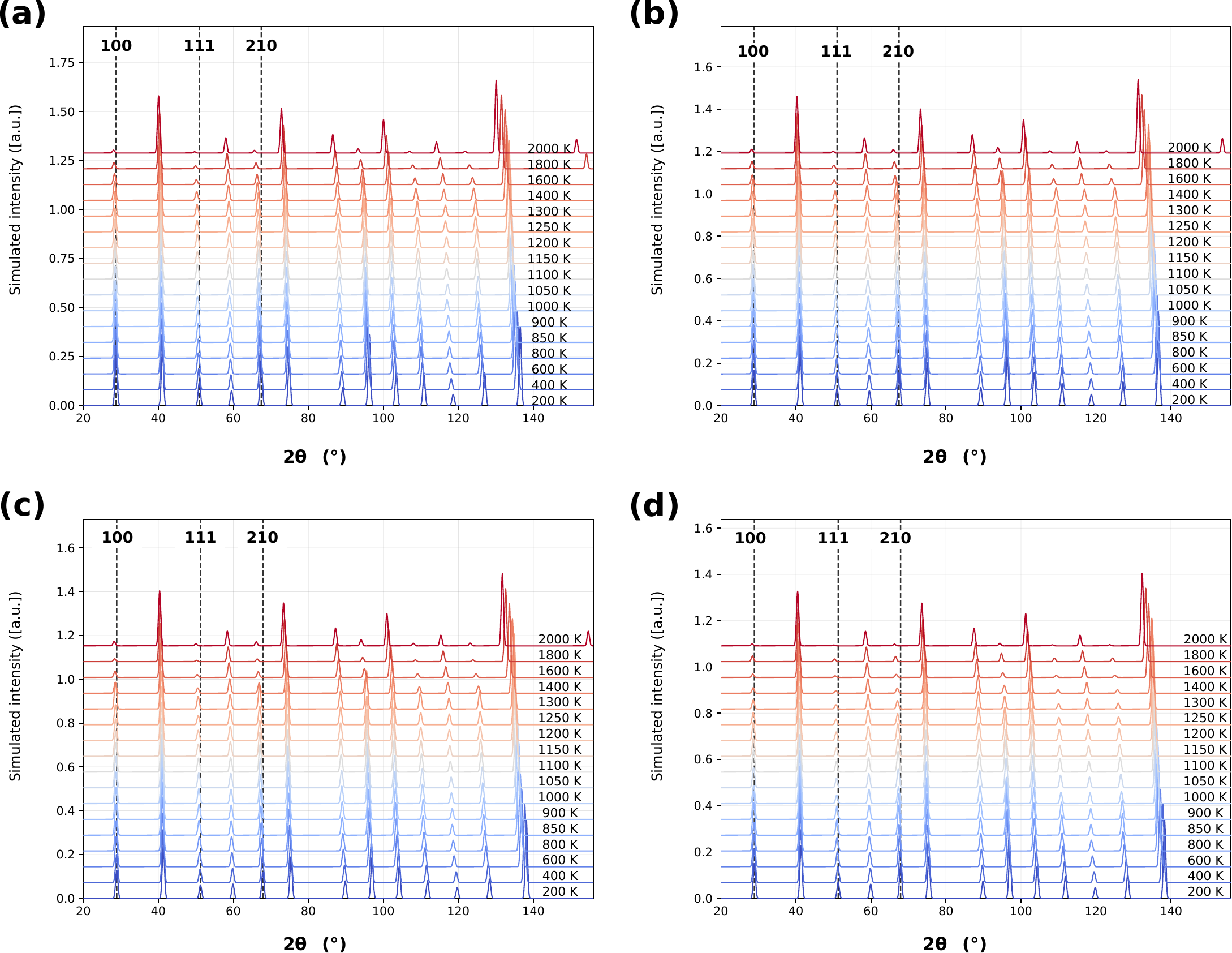}
    \caption{\add{Simulated diffraction patterns of the (a) Al25, (b) Al16, (c) Al10, and (d) Al4 alloys at selected temperatures (in Kelvin). The temperature corresponding to each pattern is indicated on the right. The vertical dashed lines indicate the positions of the characteristic B2-type superlattice reflections identified in the simulated diffraction patterns of the corresponding configurations at $200\,\text{K}$.}}
%    \textcolor{red}{Why do the superlattice peaks appear at 1800\,K for Al25 and Al4?}\\
%    \textcolor{red}{The complete lost of superlattice reflections happen at higher $T$ than what where the peaks in $C_V$ are. Why? But overall it seems they are lost at lower $T$ for decreasing Al content, which is also the trend from $C_V$, isn't it?}
    \label{fig:simulated_diffraction_intensity}
\end{figure}

\add{To further investigate B2 ordering under different temperatures, the simulated diffraction intensities are shown in Fig.~\ref{fig:simulated_diffraction_intensity}. The B2 ordering is characterized by superlattice reflections, including 100, 111, and 210 peaks. For better visibility, the superlattice peak positions are highlighted as black dashed lines. All the alloys exhibit B2 superlattice reflections (Fig.~\ref{fig:simulated_diffraction_intensity}). With increasing temperature, the intensities of these superlattice peaks gradually decrease, indicating a reduction in the degree of B2 ordering. In addition, thermal expansion causes the diffraction peaks to shift toward lower $2\theta$ degrees. The calculated 100 peak positions range from $28.745^{\circ}$ for Al25 to $28.910^{\circ}$ for Al4. These values are in good agreement with B2-related diffraction peaks reported in previous studies, located at approximately $28^{\circ}$ of NbMoCrTiAl~\cite{Laube2020-zy} and $30^{\circ}$ for MoCrTi-10Al~\cite{Chen2019-nn}.}

\add{\subsection{Short-range order parameters}}

\begin{figure}[htbp]
    \centering
    \includegraphics[width=1\linewidth]{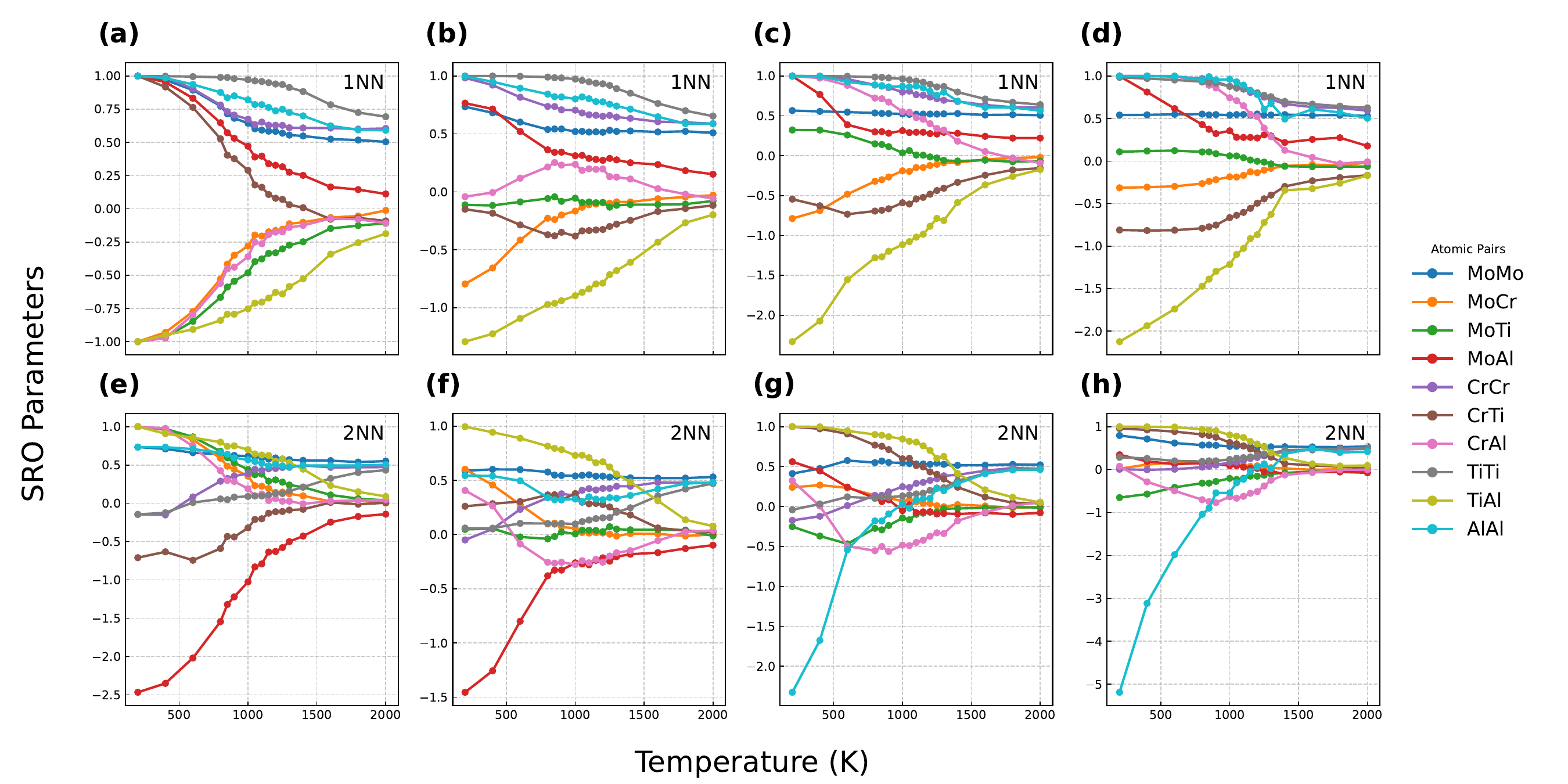}
    \caption{\add{Temperature dependence of the SRO parameters in the 1NN and 2NN shells for RHEAs: (a, e) Al25, (b, f) Al16, (c, g) Al10, and (d, h) Al4.}}
    \label{fig:SRO Parameters}
\end{figure}

To investigate the underlying mechanisms of order-disorder transition, the calculated temperature-dependent SRO parameters for all four systems are shown in Fig.~\ref{fig:SRO Parameters}. In all the first (1NN) and second (2NN) nearest-neighbor shells of the systems, a distinct trend is observed with increasing temperature: the SRO parameters for like-atom pairs (Mo--Mo, Cr--Cr, Ti--Ti, and Al--Al) tend to converge towards 0.5, while those for unlike-atom pairs (Mo--Cr, Mo--Ti, Mo--Al, Cr--Ti, Cr--Al, and Ti--Al) gradually converge towards 0. 

The evolution of SRO parameters in Fig.~\ref{fig:SRO Parameters} provides an explanation for the \add{configurational} heat capacity variations observed in Fig.~\ref{fig:Heat_capacities_of_alloys}. While Ti--Al in 1NN shell consistently remains the dominant pair for B2 ordering as indicated by its strongly negative SRO values (Figs.~\ref{fig:SRO Parameters}(a--d)), the secondary atomic pairs governing the transition behavior evolve. The variation in Al content fundamentally alters the specific atomic pairs that drive the order-disorder transition, as evidenced by the SRO parameters. In the equiatomic Al25 alloy (Figs.~\ref{fig:SRO Parameters}(a) and (e)), a synchronized change of multiple SRO parameters occurs around $900\,\text{K}$ both in 1NN and 2NN shells. Compared with the relatively weak variations at low temperatures below $500\,\text{K}$ and at high temperatures above $1500\,\text{K}$, the SRO parameters show a pronounced change in slope near $900\,\text{K}$, indicating a transition from one configuration to another. This dramatic change of Mo--Al and Cr--Ti pairs in 2NN, and Mo--Cr, Mo--Ti, Cr--Al, and Ti--Al pairs in the 1NN shell, indicates a collective breakdown of the B2 sublattice. This massive, simultaneous rearrangement of atomic bonds releases significant energy, resulting in the prominent \change{$C_v$}{$C_{V,conf}(T)$} peak observed around $900\,\text{K}$ (blue circles in Fig.~\ref{fig:Heat_capacities_of_alloys}). As the Al content decreases to 16 at.\%, the disordering events become decoupled, leading to the bimodal \change{$C_v$}{$C_{V,conf}(T)$} behavior (red squares in Fig.~\ref{fig:Heat_capacities_of_alloys}). The low-temperature \change{$C_v$}{$C_{V,conf}(T)$} peak at $\approx600\,\text{K}$ of Al16 alloy correlates strongly with the rapid evolution of the Mo--Al pair in the 2NN shell. As shown in Figs.~\ref{fig:SRO Parameters}(b) and (f), while other pairs remain relatively stable or change slowly, the Mo--Al interaction undergoes a significant transformation in a narrow temperature range, acting as the primary trigger for the low-temperature peak. At the same time, the slow transition of other unlike-atom pairs in the 1NN shell, and in particular the Ti-Al pair in both 1NN and 2NN shells, leads to a high temperature peak. As the Al content decreases to 10 at.\%, the sharp low-temperature peak at $400\,\text{K}$ (orange triangles in Fig.~\ref{fig:Heat_capacities_of_alloys}) appears. As shown in Figs.~\ref{fig:SRO Parameters}(c) and (g), the Al-Al pair in 2NN shell shows an abrupt transition from around $-2.5$ to $-0.5$ at low temperatures below $\approx600\,\text{K}$ and slowly reaches 0.5 for higher temperatures. This suggests that the rearrangement of Al--Al correlations (strong clustering on the same sublattice) is the specific reason for the sharp low-temperature peak in this composition, whereas the slow transition of other pairs in the 1NN shell leads to the high temperature peak. When the Al content decreases to 4 at.\%, the low-temperature peak vanishes (purple rhombus in Fig.~\ref{fig:Heat_capacities_of_alloys}). Instead, a synchronized evolution of all the pairs in both shells at high temperatures ($>1200\,\text{K}$) is observed (Figs.~\ref{fig:SRO Parameters}(d) and (h)). This synchronized high-temperature disordering gives rise to the single, sharp \change{$C_v$}{$C_{V,conf}(T)$} peak observed at $1250\,\text{K}$.

\begin{figure}[htbp]
    \centering
    \includegraphics[width=1\linewidth]{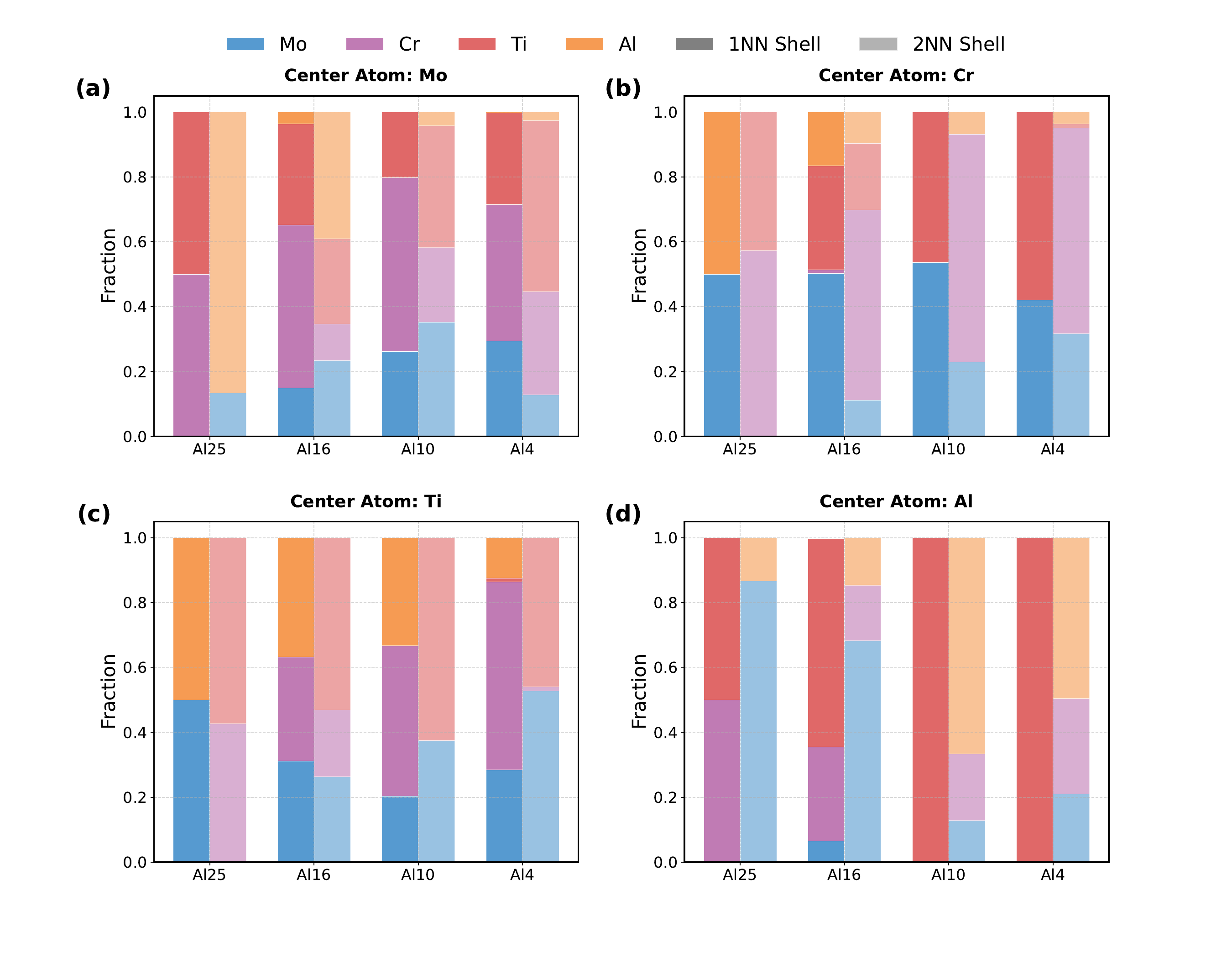}
    \caption{Average atomic fractions in the 1NN and 2NN shells around center atoms of (a) Mo, (b) Cr, (c) Ti, and (d) Al in ordered configurations at $200\,\text{K}$. The left (fully opaque) bars represent the 1NN shell, while the right (semi-transparent) bars correspond to the 2NN shell.}
    \label{fig:element fractions of 1NN and 2NN}
\end{figure}

To further investigate the underlying mechanism of element pairing, the statistical analysis of element fractions in the 1NN and 2NN shells of the ordered configuration at $200\,\text{K}$ is presented in Fig.~\ref{fig:element fractions of 1NN and 2NN}. The results reveal a distinct sublattice separation behavior driven by specific atomic preferences. Firstly, analyzing the direct neighborhood of Mo and Cr atoms, as shown in Figs.~\ref{fig:element fractions of 1NN and 2NN}(a) and (b), reveals a strong preference to occupy each other's 1NN shells, forming Mo--Cr pairs. Conversely, the local environments of Ti and Al (Figs.~\ref{fig:element fractions of 1NN and 2NN}(c) and (d)) demonstrate a pronounced tendency for chemical ordering. As shown in Fig.~\ref{fig:element fractions of 1NN and 2NN}(c), Al consistently tends to maximally occupy the 1NN shell sublattice of Ti, resulting in a local Al/(Mo+Cr) ratio that significantly exceeds the nominal macroscopic ratio when Ti is excluded.

More significantly, Fig.~\ref{fig:element fractions of 1NN and 2NN}(d) reveals that as the Al content decreases, the fraction of Ti atoms in the Al 1NN shell increases dramatically to 100\%, indicating the formation of B2 Ti--Al atomic pairs. This ordering mechanism is governed by the interplay of chemical and structural factors: thermodynamically, the highly negative formation enthalpy of the Ti--Al B2 phase ($\approx-0.3\,\text{eV/atom}$~\cite{Dehghani2022-yt}) provides a strong driving force for Ti--Al coordination \add{compared to other atomic pairs~\cite{Takeuchi2005-td}}. Structurally, the atomic-size mismatch among Al, Mo, Cr, and Ti gives rise to local lattice distortion, which may further modify the local strain energy and thereby contribute to the stabilization of chemically ordered Ti--Al pairs. \add{The atomic-size mismatch $\delta$ is calculated as~\cite{Zhang2008-ng}:
\begin{equation}
\delta =\sqrt{\sum_{i=1}^{N}c_i(1-r_i/r_{avg})^2}\ ,
\label{eq:atomic_size_misfit}
\end{equation}
where $N$ is the number of constituent elements, $c_i$ and $r_i$ are the atomic fraction and atomic radius of element $i$, respectively. $r_{\mathrm{avg}}=\sum_{i=1}^{N}c_i r_i$ is the average atomic radius. The atomic radii were taken from Ref.~\cite{Takeuchi2005-td}.
With decreasing Al content, the atomic-size misfit raise from $6.06\%$ of Al25 alloy to $6.54\%$ of Al4 alloy which may contribute to form Ti--Al atomic pairs.} This behavior is further corroborated by the increased Al fraction in the 2NN shell as shown in Fig.~\ref{fig:element fractions of 1NN and 2NN}(d). This $\text{Al}_\text{center} \rightarrow \text{Ti}_\text{1NN}\rightarrow \text{Al}_\text{center}$ atomic arrangement is a signature of the B2 structure. These findings are in good agreement with the strong ordering tendencies observed in the SRO parameters (Figs.~\ref{fig:SRO Parameters}), where the Ti--Al pair consistently appears as the dominant ordering pair. As for the statistical results for the disordered configuration at $2000\,\text{K}$ shown in Fig.~S6, the average atomic fractions within the 1NN (opaque bars) and 2NN (semi-transparent bars) shells around any constituent element are nearly identical, indicating comparable probabilities of finding these elements in either the 1NN or 2NN shells.  These local concentrations closely match the macroscopic nominal compositions across all simulated alloys from Al25 to Al4. Such random lattice occupancy confirms the formation of a fully disordered solid solution at the high-temperature state.

Furthermore, our results are consistent with the work of Laube et al.~\cite{Laube2022-jb}, who reported significant enrichment of Ti and Al in B2-ordered precipitates relative to the A2 matrix in the (TaMoTi)$_{82}$Cr$_8$Al$_{10}$ (at.\%) alloy. \add{Increasing the Al content in TaMoTiCrAl alloys was also found to drive a structural evolution from a single A2 phase to an A2 matrix containing B2 precipitates, followed by a B2 matrix containing A2 precipitates, and ultimately to a single B2 phase~\cite{Laube2021-sn}. Nevertheless, owing to their multicomponent nature, RHEAs may exhibit competition among several crystal structures and intermetallic phases. For example, in addition to the HCP (A3), B19, A2, and B2 structures considered in the Ti--Al system~\cite{Dehghani2022-yt, Abdoshahi2021-hm}, the relevant binary subsystems contain potential competing compounds such as TiCr$_2$ in Ti--Cr~\cite{Flores2024-eg}, AlMo$_3$ and AlMo in Al--Mo~\cite{Liu2024-cl, Saunders1997-dn}, and AlCr$_2$ in Al--Cr~\cite{Milosavljevic2021-un}. Simulations of the ternary Mo--Al--Ti system have further predicted that B19 can become energetically favorable over certain composition ranges~\cite{Dehghani2022-yt, Abdoshahi2021-hm}. These structures therefore represent possible competing phases in the quaternary MoCrTiAl alloys. Experimentally, however, the reported structural evolution of MoCrTiAl alloys is predominantly associated with A2/B2 ordering, and no other phases have been identified~\cite{Laube2024-bw, Laube2020-zy} in this system. On the one hand, the high configurational entropy nature of RHEAs may favor a single-phase solid solution over alloys with intermetallic phases~\cite{Miracle2017-rr}. Meanwhile, the strong Ti--Al chemical affinity promotes Ti--Al ordering, whereas the (Mo, Cr)-rich chemical environment exhibits a single A2 phase along the whole compositional range~\cite{Jindal2013-ec}.  Although the present calculations are constrained to a prescribed BCC lattice and therefore cannot explicitly assess the stability of possible competing structures or multiphase states, including B19, HCP, and Laves phases, the predicted chemical-ordering trends within the BCC-constrained configurational space are consistent with the experimentally observed A2/B2 evolution. In particular, both indicate the important role of Al in promoting B2-type ordering and the development of pronounced Ti--Al chemical ordering.}

\subsection{B2 configurations} 

\begin{figure}[htbp]
    \centering
    \includegraphics[width=0.6\linewidth]{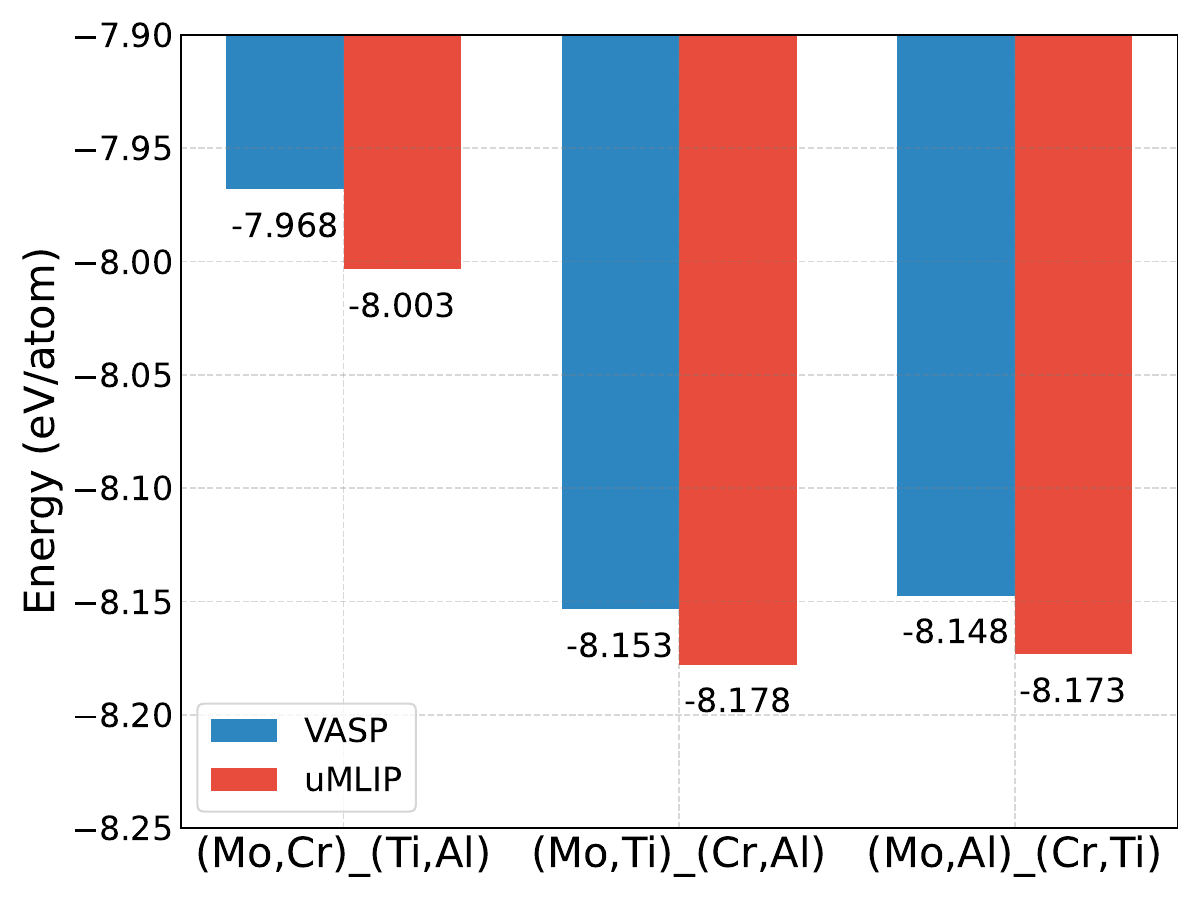}
    \caption{Potential energy of different relaxed B2 configurations at $0\,\text{K}$ from calculations via the DFT-VASP and uMLIP-GRACE methods.}
    \label{fig:calculations of B2 configurations}
\end{figure}

To investigate the competition between dominant B2 configurations during the order-disorder transition, the relaxed energies of various B2 configurations were calculated using DFT (VASP) and uMLIP (GRACE) methods at $0\,\text{K}$. As shown in Fig.~\ref{fig:calculations of B2 configurations} (for naming conventions see Table~\ref{tab:configurations_names}), the results from VASP and uMLIP exhibit remarkable consistency, with energy differences between the two methods for identical configurations remaining within $\approx0.035\,\text{eV/at.}$, which indicates the reliability of the used GRACE uMLIP.

The configurations with (Mo, Ti) sharing one sublattice and (Cr, Al) occupying the other exhibit the lowest energy (VASP: $-8.153\,\text{eV/at.}$, uMLIP: $-8.178\,\text{eV/at.}$), indicating that this atomic arrangement corresponds to the most stable state at $0\,\text{K}$. However, the configuration with (Mo, Al) sharing one sublattice and (Cr, Ti) sharing the other (VASP: $-8.148\,\text{eV/at.}$, uMLIP: $-8.173\,\text{eV/at.}$) show energy only slightly higher than the most stable state. Crucially, the energy difference between these two competing configurations is only $0.005\,\text{eV/at.}$ in both VASP and uMLIP calculations. The thermal energy scale vastly exceeds the energy difference between the two competing states in the Metropolis algorithm, where $k_BT \approx 0.026\,\text{eV/at.}$ at $300\,\text{K}$, $\approx 0.052\,\text{eV/at.}$ at $600\,\text{K}$, and $\approx 0.086\,\text{eV/at.}$ at $1000\,\text{K}$ --- respectively 5, 10, and 17 times larger than the $0.005\,\text{eV/at.}$ energy difference. Energy difference alone is therefore insufficient to dictate a unique sublattice ordering. While the (Mo, Ti)--(Cr, Al) state is energetically most preferred at $0\,\text{K}$, the MC simulations of the equiatomic Al25 system suggest a preference for the (Mo, Al)--(Cr, Ti) grouping (indicated by the positive SRO parameters of Mo-Al and Cr-Ti pairs in the 1NN shell and negative in the 2NN shell shown in Figs.~\ref{fig:SRO Parameters}(a) and (e)), implying that Mo and Al tend to co-occupy the same sublattice. This aligns with findings by Dehghani et al.~\cite{Dehghani2022-yt, Abdoshahi2021-hm}, who reported a strong preference for Mo on the Al sublattice in B2-ordered TiAlMo alloys.

\subsection{Mechanical properties}

\begin{figure}[htbp]
    \centering
    \includegraphics[width=1\linewidth]{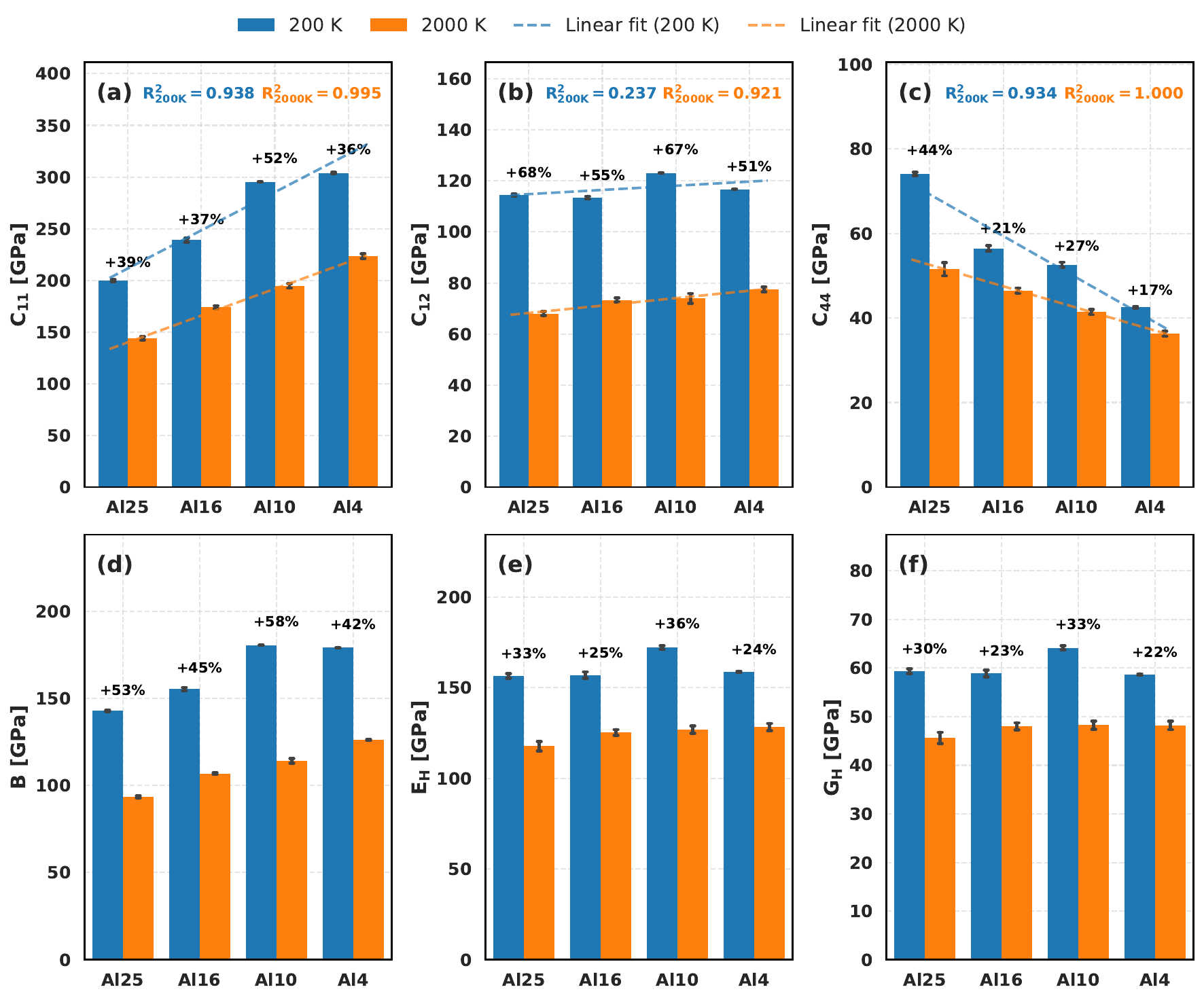}
    \caption{Mechanical properties of ordered ($200\,\text{K}$) and disordered ($2000\,\text{K}$) structures: (a--c) Elastic constants $C_{11}$, $C_{12}$, and $C_{44}$, with dashed lines indicating linear fits to the compositional dependence ($R^2$ values annotated); (d) bulk modulus $B$; (e) Young's modulus $E_H$; (f) shear modulus $G_H$. Percentage values indicate the enhancement of the ordered structure relative to the disordered counterpart.}
    \label{fig:mechanical properties}
\end{figure}

To investigate the influence of SRO on mechanical properties, we calculated the elastic constants and moduli at $0\,\text{K}$ for atomic configurations equilibrated at $200\,\text{K}$ (ordered structure) and $2000\,\text{K}$ (disordered structure). The calculated properties, including single-crystal elastic constants ($C_{11}$, $C_{12}$, and $C_{44}$), bulk modulus ($B$), polycrystalline Hill's averaged Young's modulus ($E_H$) and shear modulus ($G_H$) are summarized in Fig.~\ref{fig:mechanical properties}. Generally, alloy stiffness increases with decreasing Al content. For the disordered configurations ($2000\,\text{K}$) shown with orange bars in Figs.~\ref{fig:mechanical properties}(a--c), the predicted elastic constants demonstrate a strictly monotonic variation. Linear regression analysis confirms this strong correlation, yielding coefficients of determination ($R^2$) close to unity (i.e., $C_{11}$: $\approx0.995$ , $C_{12}$: $\approx0.921$, and $C_{44}$: $\approx1.000$). Such a trend aligns well with the rule of mixtures (ROM)~\cite{Sobol2021-cz}, which estimates the elastic constants as:
\begin{equation}
    E_{mix} = \Sigma{c_iE_i}
    \label{eq:mix rules of elastic constant} 
\end{equation}
where $E_{mix}$ represents any elastic constant of the alloy system, while $c_i$ and $E_i$ represent the atomic fraction and the corresponding elastic constant of the $i$-th constituent element, respectively. 

As illustrated by the blue bars in Figs.~\ref{fig:mechanical properties}(a--c), the ordered structures ($200\,\text{K}$) consistently exhibit enhanced stiffness compared to the disordered state ($2000\,\text{K}$). However, a significant deviation from the linear trend which is indicated by the smaller $R^2$ (i.e., $C_{11}$: $\approx0.938$, $C_{12}$: $\approx0.237$, and $C_{44}$: $\approx0.934$) is observed in the ordered configurations compared to disordered ones. Specifically, $C_{11}$ exhibits a remarkable $+52\%$ enhancement in Al10 alloy distinct from the neighboring $+37\%$ and $+36\%$ increases, while $C_{12}$ displays a local peak also at Al10 alloy. Similarly, $C_{44}$ shows anomalous strengthening in Al10, reaching a $+27\%$ increase that notably exceeds the values extrapolated from the neighboring Al16 ($+21\%$) and Al4 ($+17\%$) alloys. These variations in the fundamental constants directly translate to the moduli presented in Figs.~\ref{fig:mechanical properties}(d--f). Unlike the monotonic trend predicted by Eq.~\eqref{eq:mix rules of elastic constant}, where stiffness is expected to increase monotonically as the concentrations of the stiffer elements (i.e., Mo, Cr, and Ti) increase relative to Al, the ordered moduli exhibit a distinct trend and peak at the Al10 alloy. The strengthening effect of SRO is particularly significant in the Al10 alloy, where the ordered structure exhibits a massive improvement over the random configuration: $\approx58\%$ in bulk modulus (Fig.~\ref{fig:mechanical properties}(d)), $\approx36\%$ in Young's modulus (Fig.~\ref{fig:mechanical properties}(e)), and $\approx33\%$ in shear modulus (Fig.~\ref{fig:mechanical properties}(f)). Since thermal effects are excluded from these static calculations, the observed variations are purely structural. 

The elastic constants and moduli captured the strengthening effect from SRO in these RHEAs, which is in good agreement with experimental observations. Laube et al.~\cite{Laube2020-zy} observed that the transition from a partially ordered MoCrTi--10Al alloy to a disordered MoCrTi--5Al alloy significantly impacts both ductility and strength in the MoCrTiAl systems. Furthermore, they also found that the ordered MoCrTi--15Al alloy showed superior yield strength and hardness compared to \add{the} disordered MoTiCr--3Al alloy~\cite{Laube2024-bw}, highlighting the critical influence of the degree of ordering on mechanical properties.

\begin{figure}[htbp]
    \centering
    \includegraphics[width=1\linewidth]{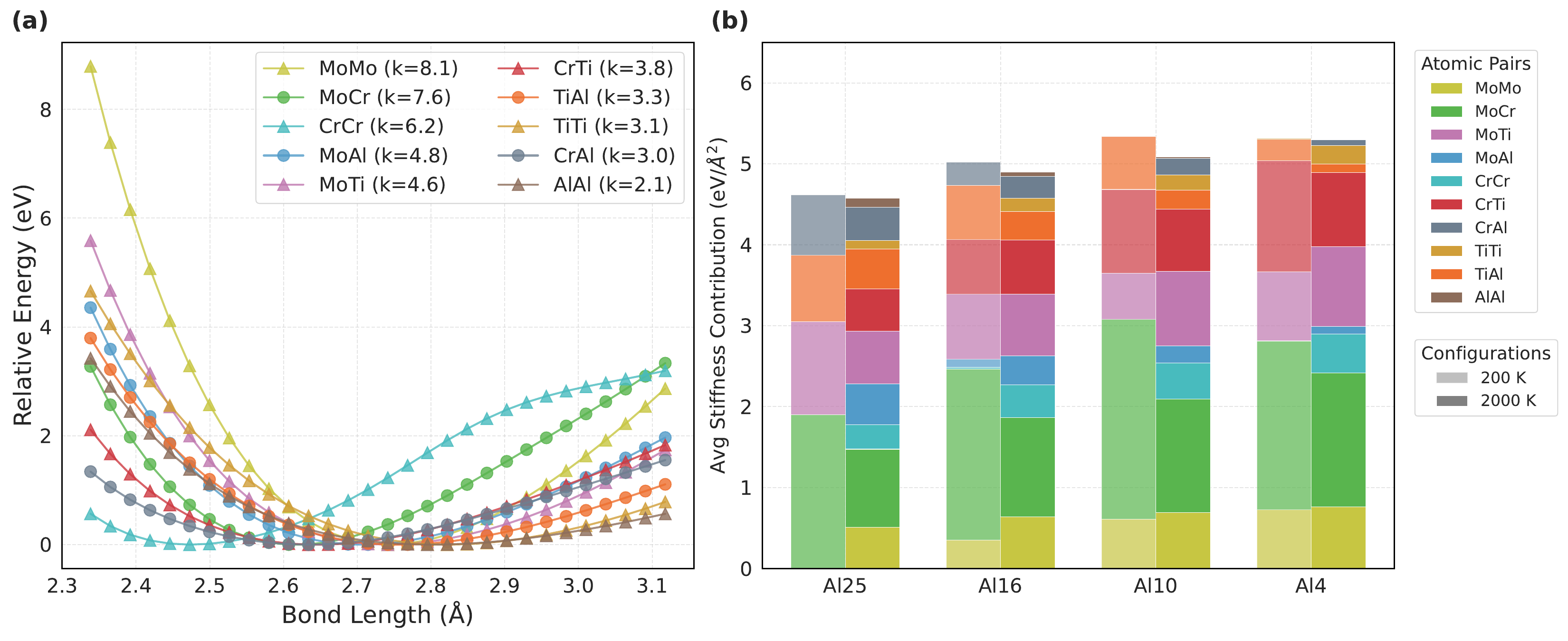}
    \caption{(a) Bond energy landscapes (with the bond stiffness $k$ provided in the legend in eV/\AA$^2$, as obtained from \add{a} parabolic fit around the respective minima) and (b) average stiffness contributions of different pairs. The left semi-transparent bars correspond to the $200\,\text{K}$ configurations, while the right bars (fully opaque) represent the $2000\,\text{K}$ configurations.}
    \label{fig:combined_stiffness_analysis}
\end{figure}

This enhancement in moduli is associated with SRO-induced changes in atomic pairing. Song et al.~\cite{Song1999-qn} established that the bulk modulus is positively correlated with the second derivative of the binding energy at equilibrium. This indicates that elastic moduli are fundamentally related to the local curvature of the energy landscape, thereby providing a physical basis for using pair stiffness as a qualitative descriptor of modulus variations. By modeling the two-atom system as a spring governed by Hooke's law, the bond stiffness $k$ can be derived from the bond energy landscapes by fitting the neighborhood of the energy minimum with a parabola for each atomic pair. Consequently, the stiffness of the alloy can be estimated using the ROM (analogous to Eq.~\eqref{eq:mix rules of elastic constant}): specifically, a higher concentration of stiff atomic pairs contributes to a larger overall modulus. Fig.~\ref{fig:combined_stiffness_analysis}(a) displays the bond energy landscapes for different atomic pairs, where relative energy is defined as the energy difference with respect to the respective ground state. The Mo--Mo pair exhibits the highest stiffness ($k = 8.1\,\text{eV/\AA}^2$), whereas the Al--Al pair is the softest ($k = 2.1\,\text{eV/\AA}^2$). 

Fig.~\ref{fig:combined_stiffness_analysis}(b) presents the relative contributions of these pairs, calculated by combining their bond stiffnesses with their fractions. In the random alloys ($2000\,\text{K}$, right fully opaque bars), global stiffness increases as the Al content decreases. However, in the ordered alloys ($200\,\text{K}$, left semi-transparent bars), variations in pair fractions result in a non-monotonic trend where stiffness first increases and then decreases. Notably, the sum of stiffness contributions peaks at Al10. \add{The pair-resolved analysis provides a microscopic explanation for the stiffness maximum at Al10. At $200\,\text{K}$, the fraction of Mo--Cr pairs in Al10 reaches $32.44\%$, which is higher than those in Al25 ($25.00\%$), Al16 ($27.77\%$), and Al4 ($27.44\%$). It is also substantially higher than the corresponding value of $18.48\%$ in the chemically disordered Al10 configuration at $2000\,\text{K}$, representing an increase of approximately $75.5\%$. Moreover, the combined fraction of Mo--Mo and Mo--Cr pairs reaches $\approx39.50\%$, the highest among the investigated compositions. Although Al4 contains more refractory elements, its Mo--Cr fraction ($\approx27.44\%$) is lower and thus shows lower cumulative stiffness.} This behavior is fully consistent with the variations in the explicitly calculated stiffness presented in Figs.~\ref{fig:mechanical properties}(a--f).

\section{Conclusions}
\change{In this work, the temperature dependent heat capacity, SRO parameters, and mechanical properties of (MoCrTi)$_{100-x}$Al$_x$ RHEAs were investigated using hybrid MC/MD simulations with a uMLIP. The calculated heat capacities and SRO parameters reveal clear order--disorder transition behavior in all studied compositions, while the transition pathway strongly depends on the Al concentration. For the Al25 and Al4 alloys, a single order--disorder transition is observed, which is dominated by the cooperative evolution of B2-type pair correlations. For Al16 and Al10 alloys, they exhibit two distinct transitions, one at low temperature and the other at higher temperature. The low-temperature transition is mainly driven by Mo--Al ordering in Al16 and Al--Al ordering in Al10, whereas the high-temperature transition is governed by the remaining atomic pairs. Structural analysis reveals that at the low temperature B2 configuration, Mo and Al share one sublattice, while Cr and Ti share the other. The mechanical response is found to be strongly coupled with the ordering state. In random configurations, stiffness increases monotonically with decreasing Al content. However, in ordered configurations, the stiffness follows a non-monotonic trend, peaking at the Al10 alloy. This behavior is associated with SRO-induced changes in atomic-pair populations, where the contribution from atomic pairs with high stiffness is maximized in the Al10 composition. These results demonstrate the important role of SRO in regulating both thermodynamic transitions and mechanical stiffness in (MoCrTi)$_{100-x}$Al$_x$ RHEAs. The present findings provide useful atomistic insights for composition optimization and the design of high-performance RHEAs.}{In this work, the temperature-dependent configurational heat capacity, chemical ordering, and mechanical properties of (MoCrTi)$_{100-x}$Al$_x$ RHEAs were investigated using hybrid MC/MD simulations combined with a uMLIP. The calculated configurational heat capacities and SRO parameters reveal composition-dependent order--disorder transition behavior. Al25 and Al4 exhibit a single dominant ordering stage associated with cooperative changes in multiple B2-type pair correlations. By contrast, Al16 and Al10 exhibit two distinct ordering stages. Their low-temperature features are primarily associated with changes in the Mo--Al and Al--Al correlations, respectively, whereas their high-temperature features involve collective changes in the remaining B2-type pair correlations. Atomic configurations, sublattice occupations, and simulated diffraction intensities consistently indicate pronounced B2-type chemical ordering at low temperatures. In the ordered configurations, Mo and Al preferentially occupy one BCC sublattice, whereas Cr and Ti preferentially occupy the other. With increasing temperature, the B2-reflection progressively decrease in intensity, indicating an evolution toward a chemically disordered state. Chemical ordering also substantially alters the composition dependence of mechanical stiffness. For the chemically disordered configurations, the elastic constants and moduli increase approximately monotonically with decreasing Al content. In contrast, the ordered configurations exhibit a non-monotonic dependence, with the highest stiffness occurring in Al10. This enhancement is associated with an SRO-induced redistribution of atomic-pair populations, which maximizes the weighted contribution of mechanically stiff pairs, particularly Mo--Cr pairs, in Al10. Overall, these results establish an atomistic connection among Al concentration, chemical-ordering pathways, and mechanical stiffness in (MoCrTi)$_{100-x}$Al$_x$ RHEAs. Although the present simulations are constrained to a prescribed BCC lattice and therefore do not explicitly assess competing crystal structures or multiphase states, they provide useful insights into chemical ordering and relative configurational stability within the BCC-constrained configuration space. These findings may facilitate the composition-guided design of RHEAs with tailored thermodynamic and mechanical properties.}

\section*{CRediT authorship contribution statement}
Jiyao Zhang: Writing---original draft, Visualization, Software, Methodology, Investigation, Formal analysis. Klemens Lechner: Writing---review \& editing, Validation. Markus Maßwohl: Writing---review \& editing, Validation. Petra Spoerk-Erdely: Writing---review \& editing, Validation. David Holec: Supervision, Writing---review \& editing, Resources, Methodology, Conceptualization.

\section*{Declaration of competing interest}
The authors declare that they have no known competing financial interests or personal relationships that could have appeared to influence the work reported in this paper.

\section*{Acknowledgments}
Jiyao Zhang acknowledges financial support from the China Scholarship Council (CSC, Grant No.202406220050). We acknowledge Austrian Scientific Computing (ASC) for awarding this project access to the LEONARDO supercomputer, owned by the EuroHPC Joint Undertaking, hosted by CINECA (Italy), and the LEONARDO consortium.

%\appendix
%\section{Example Appendix Section}

%Appendix text.

% 
% \bibliographystyle{elsarticle-num} 
% \bibliography{paperpile}

\end{document}